# Cavitation model of the initial stage of Big Bang

Mikhail N. Shneider[1], Mikhail Pekker

[1]Department of Mechanical and Aerospace Engineering, Princeton University, Princeton, NJ USA

m.n.shneider@gmail.com

**Abstract**

In this paper, we propose a model for the initial stage of the development of the Universe, analogous to cavitation in a liquid in a negative pressure field. It is assumed that at the stage of inflation, multiple breaks of the metric occur with the formation of areas of physical vacuum in which the generation of matter occurs. The proposed model explains the large-scale isotropy of the Universe without ultrafast inflationary expansion and the emergence of a large-scale cellular (cluster) structure, as a result of the development of cavitation ruptures of a false vacuum. It is shown that the cavitation model can be considered on par with (or as an alternative to) the generally accepted inflationary multiverse model of the Big Bang.

## I. Introduction

One of the main problems of the Big Bang theory is the amazing homogeneity and isotropy of the observed cosmic microwave background (CMB) radiation [1,2] discovered in 1964 by Penzias and Wilson [3] and predicted earlier by Gamow [4]. Also a mystery is the cluster structure of the distribution of visible matter in the universe and its uniformity at sizes exceeding 100 mega parsecs [1,2]. The inflationary model of the initial stage of the expansion of the universe, proposed by Guth [5] and further developed by Linde [6-10], Vilenkin [11-14] and others [15-28], explains the observed isotropy and uniformity of the CMB, but causes a mixed reaction in the scientific the world, since it assumes the generation of $\sim 10^{80}$ unrelated space-time universes at the stage of inflationary expansion. The existence of our universe (one of the myriad universes) where, as a result of evolution, life of the terrestrial type arose, is explained by the anthropic principle. Thus, to be consistent with the inflationary model, the Hubble constant, characterizing the expansion of our Universe, varied from $H \sim 10^{42} - 10^{36}$ $s^{-1}$, to the present value $H \approx 2.2 \cdot 10^{-18}$ $s^{-1}$ [29-34].

The inflationary theory of the expansion of the universe is based on the presence of the Λ-term in the equations of the general theory of relativity. The acceleration of the expansion of the universe recently discovered by Perlmutter, Riess, and Schmidt [35-38], is also explained by the presence of the Λ-term in the equations of the dynamics of the universe. Nevertheless, many cosmologists are skeptical about the existence of a superfast inflationary stage of the Big Bang [29-34].

In this paper, a cavitation model of the inflationary stage of the big bang is proposed, which is similar to cavitation in a liquid in a negative pressure field. In this model, the effect of the cosmological Λ-term is considered as a source of negative pressure acting in the region of a false vacuum, and the physical vacuum appears as vacuum bubbles arising due to the breaking of the metric. The cavitation model makes it possible to explain the homogeneity and large-scale isotropy of the universe as well as the large-scale cellular structure of the universe without generating unrelated space-time universes.

The introductory Section II briefly considers: a) the phenomenon of cavitation in a liquid and the ideas underlying its understanding, b) the legitimacy of the application of these ideas to the early stage of the expansion of the universe, and c) the difference between the standard inflationary and cavitation models in their description of the transition from a false vacuum to a physical one.

Section III considers the proposed cavitation model of the inflationary stage of the expansion of the universe. It is shown that within the framework of this model it is possible to explain the isotropy and homogeneity of the cosmic microwave background (CMB), the distribution of visible matter, as well as the observed large-scale cellular (cluster) structure of the universe.

Section IV qualitatively examines the expansion of the "cavitation" bubble in a false vacuum and the generation of matter/energy at its borders. Qualitative considerations are presented to explain the inhomogeneous distribution of energy in cells associated with bubbles of the physical vacuum.

In the conclusion, the main results and consequences of the cavitation model for the inflationary stage of the Universe expansion are briefly formulated.

## II. Cavitation in a liquid (brief introduction)

The similarity of the observed large-scale cellular pattern of the Universe to the region of developed cavitation in a liquid is not accidental in our opinion. It is supported by the fact that at the early inflationary stage of the expansion of the universe is associated with the cosmological term in the Einstein equations, which plays the role of negative pressure in the cavitating fluid. We believe that the inflationary stage of the development of the universe is accompanied by breaks in the metric – the appearance of "bubbles" of the physical vacuum, similar to the formation of bubbles in a liquid at negative pressure, the absolute value of which exceeds a certain critical value.

The main difference between our approach and the one adopted in the inflationary theory is that the transition from a false vacuum to a physical one does not take place all at once in the entire universe due to the "rolling" of the universe into a potential well [2], but rather takes place initially at certain points which later become the cells of the large-scale structure observed in the universe. The emergence of "bubbles" of the physical vacuum significantly slows down the inflation of the universe at the inflation stage and serves as a place for the generation of matter in them. Moreover, the formation of metric discontinuities during the transition from a false vacuum (where repulsion due to the Λ-term acts) to a physical vacuum is an isotropizing factor, since negative pressure is kept in areas of false vacuum near the threshold of generation of discontinuities: any local excess of negative pressure is damped by an increase in the number of "bubbles" of the physical vacuum. Therefore, superfast inflation in times $<10^{-32}$ sec is not required for large-scale homogeneity and isotropy of the matter density distribution and CMB in our universe. Note that a similar feedback mechanism of formed cavitation bubbles, which limits the magnitude of the negative pressure in the cavitating liquid, was considered by authors in [39, 40].

For an ideal gas, the pressure is always positive -- only the pressure difference can be negative. In contrast, "negative pressure" is a key concept in fluid mechanics of liquids with stretching tension that may lead to cavitation or formation of voids. In other words, negative pressure is possible only in media, inside of which natural boundaries with a thickness much less than the characteristic size of cavities (bubbles) can exist. In a gas such "natural" boundaries, and hence negative pressure, cannot exist since gas molecules fill any empty region of space.

Thus, the presence of negative pressure at the inflationary expansion stage of the universe, caused by the presence of the Λ-term, allows us to consider the universe as a liquid in which natural boundaries can exist that separate the "false vacuum" with the equation of state $\rho + \frac{p}{c^2} = 0$ from the real physical vacuum with zero energy density or filled with matter with $p \geq 0$. The presence of boundaries between "false vacuum" and real vacuum, as is well known (see, for example, [1]), leads to the generation of particles due to tidal forces associated with the gradient of the gravitational field at the discontinuity of the metric. Below, in section IV, we will qualitatively consider these phenomena.

Three fundamental ideas lie in the theory of the formation and dynamics of cavitation bubbles in a liquid. The first concerns the cause of bubble expansion in a liquid [41], the second concerns the mechanism of bubble formation [42], and the third concerns the mechanism limiting the rate of bubble formation (the inverse effect of bubbles on their generation) [39, 40].

1. In 1917, Rayleigh introduced the concept of negative pressure and wrote a simple condition under which a bubble either expands or collapses [41]. The sum of pressures acting on the boundary surface of the bubble is

$$P = P_- + P_L, \qquad (1)$$

where $P_L = 2\sigma/R_b$ is the Laplace pressure, $\sigma$ is the surface tension coefficient, $R_b$ is the bubble radius, and $P_-$ is the negative pressure (Fig. 1). Since, with an increase in the bubble radius, the Laplace pressure drops as $\sim 1/R_b$, then for

$$|P_-| > \frac{2\sigma}{R_b} \qquad (2)$$

bubbles grow indefinitely. The source of negative pressure in a liquid can be the pressure of saturated vapors in a bubble, injection of a gas into a liquid, or stretching of the liquid itself (e.g., behind the front of a shock wave).

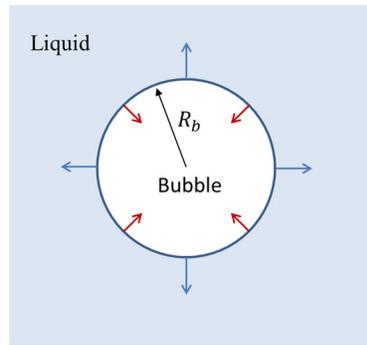

**Fig.1**. Rayleigh bubble in a negative pressure field. Red arrows show the forces of surface tension of the liquid (Laplace pressure $P_L = 2\sigma/R_b$), while the blue arrows show negative pressure $P_-$ directed outward.

2. In 1942 Zel'dovich [42] and, independently Fisher in 1948 [43], considered the formation of a bubble as a local liquid-saturated vapor phase transition under negative pressure caused by the stretching of the liquid.

According to Zel'dovich, the number of bubbles of radius $R_b$ produced per unit time depends exponentially on the energy required to create it:

$$\frac{dn_b}{dt} = \Gamma_0 exp\left(-\frac{W_b}{k_B T}\right), \qquad (3)$$

$$W_b = \int_0^{R_b} 4\pi \left(|P_-| - \frac{2\sigma}{R}\right) R^2 dR = -\frac{4\pi}{3}|P_-|R_b^3 + 4\pi R_b^2 \sigma, \qquad (4)$$

where $n_b$ is the density of bubbles of radius $R_b$ per unit volume, $k_B$ is the Boltzmann constant, $T$ is the temperature of the liquid, $W_b$ is the energy required to create a bubble of radius $R_b$, and $\Gamma_0$ is the kinetic prefactor, which depends on the theoretical model used. For example, in [39,40,44,45]

$$\Gamma_0 = \frac{3}{4\pi R_b^3} \cdot \frac{k_B T}{h}, \qquad (5)$$

where $h$ is the Planck constant.

Figure 2 shows the dependence of the energy $W_b$ on the bubble radius. The maximum of $W_b$ is reached at $R_{b,cr} = \frac{2\sigma}{|P_-|}$ and is equal to $W_{b,cr} = W_b(R_{cr}) = \frac{16\pi \sigma^3}{3|P_{cr}|^2}$. Obviously, if $R_b > R_{b,cr}$, then, in accordance with (1), the radius of the emerging bubble grows infinitely. Substituting $W_{b,cr}$ in (2), we obtain an estimate of the negative pressure at which cavitation begins:

$$|P_{cr}| \sim \left(\frac{16\pi \sigma^3}{3 k_B T}\right)^{1/2}. \qquad (6)$$

For water, as an example, $\sigma = 0.072$ N/m and $T = 300$ K, $|P_{cr}| \sim 700$ MPa. This value is more than 10 times higher than that observed in experiments [44]. Taking into account the pre-exponential factor in (5) reduces $|P_{cr}|$ to 220 MPa [3]. It was shown in [39] that considering the dependences of the surface tension coefficient on negative pressure and bubble radius makes it possible to bring the critical pressure at which cavitation begins approaches the experimentally measured values which are close to 30-50 MPa [44].

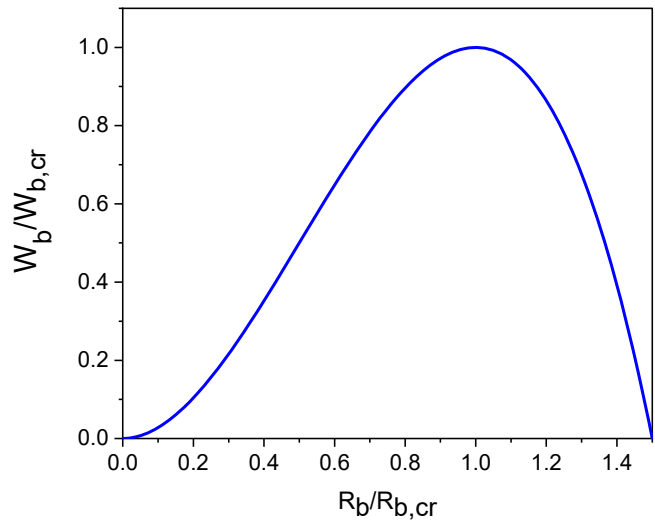

**Fig. 2.** The energy $W_b$ required to create a bubble of radius $R_b$.

Due to the exponential dependence of $\Gamma$ on $\frac{|P_-|}{k_B T}$ it follows that even with a slight excess of the absolute value of the negative pressure in the liquid above the threshold, the number of bubbles of a critical size arising in a unit volume can exceed the number of water molecules. The mechanism limiting the rate of growth of the density of cavitation bubbles was considered by the authors in [39, 40]. It was shown that for any initial value of negative pressure exceeding the critical one, it will always decrease to a value close to the critical one as a result of the expenditure of energy stored in the excess negative pressure on the generation and growth of cavitation bubbles. Below we will briefly discuss this mechanism using water as an example, since it is key to the cavitation model of the expansion of the universe at an early stage.

### II.1. Classic cavitation; saturation effect on the bubble generation rate.

Assume that the volume in which a super-critical negative pressure occurs is of the order of $l_-^3$. The formation of bubbles is associated with an increased volume of the cavitation region, which in turn leads to excessive positive pressure that reduces the value of the negative pressure in the area. The density of bubbles $n_b$ that appear in the negative pressure region $l_-^3$ over the characteristic time of pressure equilibration $t_- = \frac{l_-}{c_s}$ ($c_s$ is sound velocity) is equal to:

$$n_b = \int_0^{t_-} \Gamma dt \tag{6}$$

Accordingly, the relative change in the volume of liquid in the area where bubbles are emerging is:

$$\frac{\delta V}{V} \approx \frac{V_{cr} N_b}{l_-^3} = V_{cr} n_b \tag{7}$$

Where $V_{cr} = \frac{4}{3}\pi R_{cr}^3$ is the volume of a bubble of the critical radius $R_{cr}$ ($R_{cr}$ corresponds to the initial negative pressure $P_{-,0}$), and $n_b$ is the number of bubbles in the volume $l_-^3$.

The value of excess pressure $\delta P$ can be estimated from the simplest equation of state for a compressible fluid with sound velocity $c_s$:

$$\delta P = c_s^2 \delta\rho = c_s^2 \rho \frac{\delta V}{V} \approx V_{cr} n_b c_s^2 \rho \tag{8}$$

where $\rho = 10^3 \ kg/m^3$ is the unperturbed density of water.

The absolute value of the total pressure in the bubble generation region is equal to

$$|P_-| = |P_{-,0}| - \delta P = |P_0| - V_{cr} n_b c_s^2 \rho . \tag{9}$$

Substituting $|P_-|$ into the first equation (4) at $R_b = R_{cr}$, we get:

$$W_{cr} = W_{b,cr} + V_{cr}^2 n_b c_s^2 \rho, \tag{10}$$

Recall that $W_{cr,b}$ is the energy required to create a bubble of critical size without taking into account the stretching of liquid.

From (10), one can see that the increase in the number of cavitation voids per unit volume increases $W_{cr}$, and thus, reduces the rate of pore formation in (3).

Substituting (10) into (3) we get:

$$\frac{dn_b}{dt} = \frac{k_B T}{V_{cr} h} \exp\left(-\frac{W_{b,cr}}{k_B T}\right) \cdot exp\left(-\frac{V_{cr}^2 c_s^2 \rho}{k_B T} n_b\right) = \Gamma(W_{b,cr}) \exp\left(-\frac{V_{cr}^2 c_s^2 \rho}{k_B T} n_b\right) \quad (11)$$

The equation (11) has a solution at fixed $P_{-,0}$.

$$n_{b,satur} = \frac{k_B T}{V_{cr}^2 c_s^2 \rho_0} \ln\left(1 + \frac{V_{cr}^2 c_s^2 \rho_0}{k_B T} \Gamma(W_{b,cr}) t\right) \quad (12)$$

Figure 3 shows the dependence of $n_b(t)$ with and without the saturation effect. For small period of time when:

$$\frac{V_{cr}^2 c_s^2 \rho}{k_B T} \Gamma(W_{b,cr}) t < 1 \quad (13)$$

The density of bubbles increases linearly with time. Then, taking into account the saturation, the dependence $n_b(t)$ becomes logarithmic. Note that taking into account the dependence of the surface tension coefficient on negative pressure and bubble radius does not qualitatively change the graph shown in Fig. 3 [3, 39].

It follows from (4) that $|P_{cr}| \sim \frac{1}{\sqrt{T}}$, and at $T \to 0$, $|P_{cr}| \to \infty$. However, as was shown in experiments with liquid helium (see, for example, [45]) at low temperatures $|P_{cr}|$ does not depend on temperature.

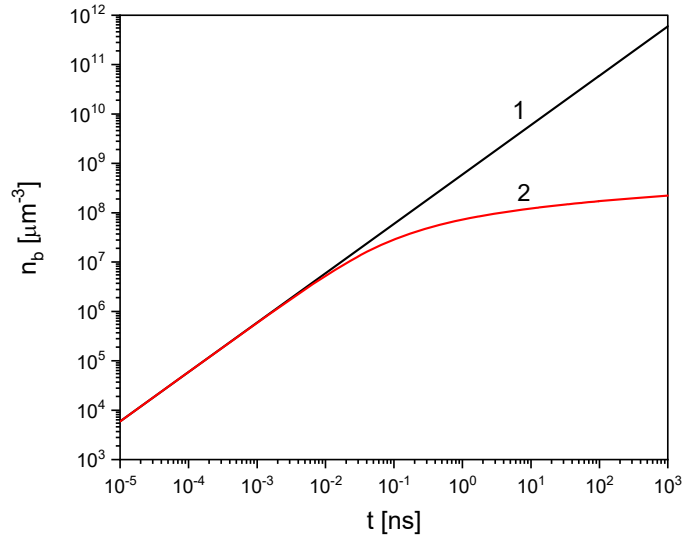

**Fig.3**. The dependence of the emerging bubbles' density on time. Line 1 – without considering the effect of saturation; line 2 – with saturation. Negative pressure $P_{cr} = -350$ MPa is obtained by taking into account the pre-exponential factor (5).

The temperature dependence of the nucleation rate (3) is called Arrhenius. Arrhenius introduced such temperature dependence for the rates of chemical reactions. If we follow the Arrhenius law, the probability of formation of cavitation bubbles in cryogenic liquids such as $^4$He and $^3$He should tend to zero at $T \to 0$. However, in experiments [45], nucleation at temperatures below 1 K did not depend on temperature, because in cryogenic liquids the formation of cavitation bubbles does not follow the Arrhenius law, but is determined by the tunnel effect.

## II.2. Cavitation in a cryogenic liquid: the tunnel effect.

The possibility of tunnel nucleation (creation of growing bubbles) in liquid helium was first considered by Lifshitz and Kagan in 1972 [47] in the quasi-classical approximation. The expression for $\Gamma$ in this case does not depend on temperature as in (3) and (4) but is determined by the eigenvalues of the energy of zero-point vibrations in the potential well and the height of the potential barrier. A theory describing the tunneling production of bubbles in the general case was developed in [39]. The approach described below can serve as a model for constructing a theory of the formation of "bubbles" of a real vacuum in a false vacuum at the stage of inflationary expansion of the universe.

The kinetic energy associated with the expansion of the bubble (kinetic energy of the added mass) is [41]:

$$E_{kin} = 2\pi R_b^3 \rho \left(\frac{dR_b}{dt}\right)^2. \tag{14}$$

In accordance with [47], we formally write the Lagrange function for a bubble as the difference between the kinetic energy of the bubble (14) and the potential energy (4):

$$L(R_b, \dot{R}_b) = E_{kin} - W_b = 2\pi R_b^3 \rho \dot{R}_b^2 - 4\pi \sigma R_b^2 + \frac{4\pi}{3}|P_-|R_b^3. \tag{15}$$

In (15), we neglected the fluid expansion (the dependence of $\rho$ on $|P_-|$) and the dependence of $\sigma$ on the bubble radius $R_b$ and $|P_-|$.

Considering that in our thin layer approximation, the velocity of the added mass $\sim \partial R_b/\partial t \equiv \dot{R}_b$, from (15) we find the canonical momentum $p_b$:

$$p_b = -\frac{\partial L}{\partial \dot{R}_b} = -4\pi \rho R_b^3 \dot{R}_b, \tag{16}$$

and the Hamiltonian corresponding to Lagrangian (15) is:

$$H_b(R_b, P_b) = \frac{p_b^2}{2M} + W_b = \frac{4\pi \rho R_b^3}{2}\dot{R}_b^2 + 4\pi \sigma R_b^2 - \frac{4\pi}{3}|P_-|R_b^3. \tag{17}$$

This Hamiltonian (17) coincides with that obtained in [47], where the equation (17) was considered in the framework of Bohr's approach in calculating the energy of zero-point oscillations.

The Hamiltonian in (17) can be regarded as the Hamiltonian of a particle with a mass $M$ depending on its coordinate $R_b$ ($M = 4\pi \rho R_b^3$), located in a given potential $W_b$. By analogy with quantum mechanics, when the Hamiltonian (17) for a point particle corresponds to the stationary Schrödinger equation, we can assume that equation (17) corresponds to the equation [39]:

$$\frac{1}{R_b^3}\frac{d}{dR_b}R_b^3\frac{d\Psi}{dR_b} + \left(\frac{2M}{\hbar^2}(E - W_b)\right)\Psi = \frac{1}{R_b^3}\frac{d}{dR_b}R_b^3\frac{d\Psi}{dR_b} + \left(\frac{8\pi \rho R_b^3}{\hbar^2}\left(E - 4\pi \sigma R_b^2 + \frac{4\pi}{3}|P_-|R_b^3\right)\right)\Psi = 0, \tag{18}$$

where $\Psi$ is the wave function. Equation (18) differs from the usual Schrödinger equation in that the particle is considered to be pointwise in the Schrödinger equation, and the square of the modulus of the wave function describes the probability of being in a given point in space. Whereas in our case, described by equation (18), "particles" correspond to a hollow sphere with an infinitely thin shell with a mass $M = 4\pi\rho R_b^3$ depending on its radius, and the square of the modulus of the wave function describes the probability that the shell will have a radius $R_b$. Of course, if the mass $M$ is constant, equation (18) goes into the usual Schrödinger equation. Thus, as for the conventional Schrödinger equation, our task is to find the spectrum of the eigenvalues $E$ and the corresponding eigenfunctions for equation (18) at a given fixed value of negative pressure $P_-$.

As an example, let's consider the tunneling regime of cavitation in $^4$He. Fig. 4 shows the potential barrier and wave functions $\chi = x\Psi_k$ corresponding to the self-energies of Eq. (18) describing tunneling nucleation for $^4$He in a negative pressure field [39]. The vertical lines on figures 4 (b) and 4 (c) correspond to the values of $x$ at which $U_b = E$. At $x_2 > x > x_1$ and the wave function decays. At $x > x_2$ the pore radius begins to grow since the surface tension cannot compensate for the tensile forces associated with negative pressure. For $T \to 0$, only the zero mode contributes to the bubble formation probability $\Gamma$. At a finite temperature of helium, higher modes also contribute. The estimate of the nucleation rate based on the quantum mechanical description [39] has the form:

$$\Gamma = \Sigma_k \frac{k_B E_k}{\hbar V_{He}} \frac{|\Psi_k(x_{2,k})|^2}{\int_0^{x_{2,k}} |\Psi_k|^2 x^2 dx} e^{-(E_k-E_0)/T} \tag{19}$$

Here $k_B$ – Boltzmann constant, $T$ – temperature of liquid, $k$ – mode number (see Fig. 4), $V_{He} = \frac{4}{3}\pi d_m^3$ – volume of Helium atom, $d_m$ – radius, $E_k$ – eigenvalue of $k$-th mode in Kelvin, $\Psi_k(x_{2,k})$ is the value of k-th wave function at point $x_2$ (see Fig. 4). It should be noted that at $|P_-| = 1.78$ MPa in $^4$He there are only two cavitation modes.

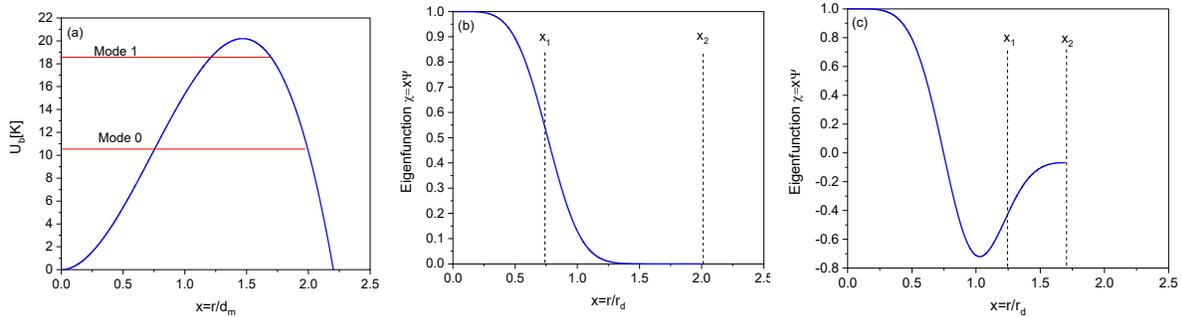

**Fig. 4.** (a) – Dependence of the potential $U_b = 4\pi\sigma R_b^2 - \frac{4\pi}{3}|P_-|R_b^3$ on the "coordinate" $x$ for $^4$He at $|P_-| = 1.78$ MPa. The horizontal lines show the eigenmodes. (b), (c) are the eigenfunctions $\chi = x\Psi$ of equation (18). The interval $(x_1, x_2)$ correspond to the attenuation of the wave functions. The energy of the zero mode (in Kelvin degrees) $E_0 = 10.8$ K, the first $E_1 = 18.6$ K. The shown results correspond to $^4$He at T<0.5 K: $\rho = 145$ kg/m$^{-3}$, $\sigma = 0.374 \cdot 10^{-3}$ Nm$^{-1}$ [48], $d_m = 0.287$ nm [39].

Below, we will study how the inflationary stage of the expansion of the universe can change, if the tunnel inception of cavitation is possible, similar to cavitation in cryogenic liquid helium.

## III. Cavitation inflation model

In traditional inflationary cosmological models, it is assumed that at the initial stage there is a phase transition from a false vacuum to a physical one, with an explosive release of energy. As a result, a superfast (inflationary) expansion of the universe occurs, which is subsequently replaced by inertial expansion [1,2]. This approach is generally accepted for most modern cosmological models with the cosmological $\Lambda$ − term [1,2]. Inflationary expansion, as we said above, was introduced to explain the isotropy of the CMB and the mass distribution on scales of the order of 1000 mega parsecs.

The equation describing the homogeneous and isotropic expansion of the universe in Planck units of time $t_p = (Gh/c^5)^{1/2} \approx 10^{-43}$ [s], ($G$ is gravitational constant, $c$ is speed of light), length $l_p = (Gh/c^3)^{1/2} \approx 10^{-35}$ [m], and density $\rho_p = (G^2 h c^{-5})^{-1} = 0.8 \cdot 10^{96}$ [kg/m³], has the form [1,2]:

$$\frac{1}{a}\frac{d^2 a}{d\tau^2} = -\frac{4\pi}{3}(\tilde{\rho} + 3\tilde{P}), \tag{20}$$

where $\tau = t/t_p$, $a = l/l_p$, $\tilde{\rho} = \rho/\rho_p$, $\tilde{P} = P/(\rho_p c^2)$.

In the case of false vacuum, $\tilde{\rho} = \tilde{\rho}_V$, $\tilde{P} = -\tilde{\rho}_V$, equation (20) takes the form [1,2]:

$$\frac{1}{a}\frac{d^2 a}{d\tau^2} = \frac{8\pi}{3}\tilde{\rho}_V = \tilde{\Lambda}. \tag{21}$$

Where $\tilde{\Lambda} = \Lambda c^2 t_p^2$, and $\Lambda$ is the cosmological term introduced by Einstein in the equations of the General Theory of Relativity (see, for example, [1]). The inflationary expansion $a = a_0 e^{\gamma_V \tau}$, where $a_0$ – initial size of the universe, $\gamma_V = \left(\frac{8\pi}{3}\tilde{\rho}_V\right)^{1/2} = \tilde{\Lambda}^{1/2}$, corresponds to the solution of equation (20).

Following the analogy with the tunneling mechanism of cavitation generation in cryogenic liquids [39], described in the previous section, we will consider space-time as a kind of quantum liquid (similar to cryogenic Helium) that can be either in a state of false vacuum or physical vacuum. Both states are separated from each other by a potential barrier, the height of which depends on the value of the cosmological $\Lambda$ −term (which determines the effective negative pressure). The higher the absolute negative pressure, the lower the height of the barrier is. We will also assume that the false vacuum is in the ground state, similar to the one considered above (the zero mode for ⁴He, shown in Fig. 4) as an example.

By analogy with cavitation in a liquid, described in Section II, we will assume that for a value of negative pressure $|\tilde{P}_-| > |\tilde{P}_{cr}|$ in the false vacuum (the region where the energy-momentum tensor in the Einstein's equations corresponds to the presence of the $\Lambda$ −term), discontinuities of the metric begin, and the "bubbles" of the physical vacuum appear, in which the energy-momentum tensor is equal to zero. If $|\tilde{P}_-| < |\tilde{P}_{cr}|$, then the probability of cavitation rupture formation is sharply reduced. The appearance of "bubbles" of the physical vacuum leads to a decrease in the absolute value of negative pressure ($\Lambda$-term), and accordingly, to a decrease in the probability of the formation of new discontinuities in the metric. Thus, space-time "cavitation" (the formation of "bubbles" of the physical vacuum) is an isotropization factor in the universe, since it keeps negative pressure in any part of the universe near the threshold for generation of discontinuities.

Since $\Lambda = 0$ in the "bubbles" of the physical vacuum, they do not contribute to the expansion of the universe. Therefore, the effective (averaged over the entire volume of the universe) negative pressure can be written as $\tilde{P}_- = -\tilde{\rho}_V(1 - \tilde{v})$, where $\tilde{v}$ is the relative volume occupied by the "bubbles" of the physical vacuum. Accordingly, equation (21) takes the form:

$$\frac{1}{a}\frac{d^2 a}{d\tau^2} = \frac{8\pi}{3} \tilde{\rho}_V (1 - \tilde{v}). \tag{22}$$

Since, with an increase in the volume of physical vacuum $\tilde{v}$, the absolute value of negative pressure $|\tilde{P}_-|$ falls, then a moment should come when the appearance of new discontinuities in the metric (areas of transition from a false vacuum to a physical one) will become impossible. Let us write out a phenomenological equation describing the generation of discontinuities, similar to equations (3) and (11):

$$\frac{dn}{d\tau} = \xi \cdot exp\left(-\frac{\beta}{\tilde{\rho}_V(1-\tilde{v})}\right), \tag{23}$$

where $\beta$ is a parameter characterizing the threshold at which the generation of discontinuities begins and $\xi$ is a pre-exponential factor similar to those of (3) and (19).

Taking into account the expansion of the universe, the relative volume occupied by the discontinuities has the form:

$$\tilde{v} = \frac{\int_0^\tau \frac{dn}{d\tau'}\left(r_0 \frac{a(\tau)}{a(\tau')}\right)^3 d\tau'}{a(\tau)^3} = \int_0^\tau \frac{a_0^3}{a(\tau')^3} \frac{dn}{d\tau'} \left(\frac{r_0}{a_0}\right)^3 d\tau', \tag{24}$$

where $a_0, r_0$ are the initial size of the universe and the discontinuity (physical vacuum "bubble") radius in Planck units. Since, regardless of the size of the universe $a$, the metric discontinuity that appeared at the moment of time $\tau'$ has the size $r_0$, the relative volume in (24) at this time $\tau'$ is equal to $(r_0/a(\tau'))^3$. In (24), we took into account that the physical vacuum bubble over time begins to expand at the rate of expansion of the universe at the time the bubble appears.

Substituting (23) into (24) we obtain:

$$\tilde{v} = \xi \int_0^\tau \frac{a_0^3}{a(\tau')^3} exp\left(-\frac{\beta}{\tilde{\rho}_V(1-\tilde{v}(\tau'))}\right) \left(\frac{r_0}{a_0}\right)^3 d\tau'. \tag{25}$$

Since with increasing $\tilde{v}$ the exponent factor $\frac{\beta}{\tilde{\rho}_V(1-\tilde{v})}$ in (25) grows as $\sim \frac{1}{(1-\tilde{v})}$, then, over time, the generation of bubbles will stop when $\frac{\beta}{\tilde{\rho}_V(1-\tilde{v})} \gg 1$. Hence, it follows that the parameter $\frac{\beta}{\tilde{\rho}_V}$ must always be less than unity, since if $\frac{\beta}{\tilde{\rho}_V} > 1$, the generation of bubbles will be significantly suppressed from the very beginning.

On the other hand, for $\tilde{v} \ll 1$, the size of the universe grows exponentially $a = a_0 e^{\gamma_V \tau}$. Therefore, the relative volume occupied by bubbles is:

$$\tilde{v} = \xi a_0^3 \int_0^\tau \frac{1}{a(\tau')^3} d\tau' = \xi a_0^3 \int_0^\tau \frac{1}{a(\tau')^3} d\tau' = \xi \int_0^\tau e^{-3\gamma_V \tau'} d\tau' \leq \frac{\xi}{3\gamma_V} \tag{26}$$

It follows from (26) that if $\frac{\xi}{3\gamma_V} \ll 1$, the bubble formation does not play a role. It should be noted that since for any values of $\frac{\beta}{\tilde{\rho}_V}$ and $\frac{\xi}{3\gamma_V}$ the relative volume of "bubbles" of the physical vacuum never reaches unity, the presence of discontinuities only reduces the exponent of the inflationary expansion of the universe.

To demonstrate the effect of cavitation at the inflationary stage, let us choose, for example, the dimensionless density $\tilde{\rho}_V = 0.01$, corresponding to $\gamma_V = 0.29/t_p$. Figure 5 shows the calculation results taking for example $\beta = 10^{-5}$, $\tilde{\rho}_V = 0.01$, $r_0 = 0.05$, for the values $\xi = 150$ and 50. Since for $\tau > 5$ the relative volume of "bubbles" of the physical vacuum ceases to depend on time (Fig. 5,a), the radius of the universe enters the exponential stage. For $\tau \geq 5$, new discontinuities do not affect the expansion of the universe.

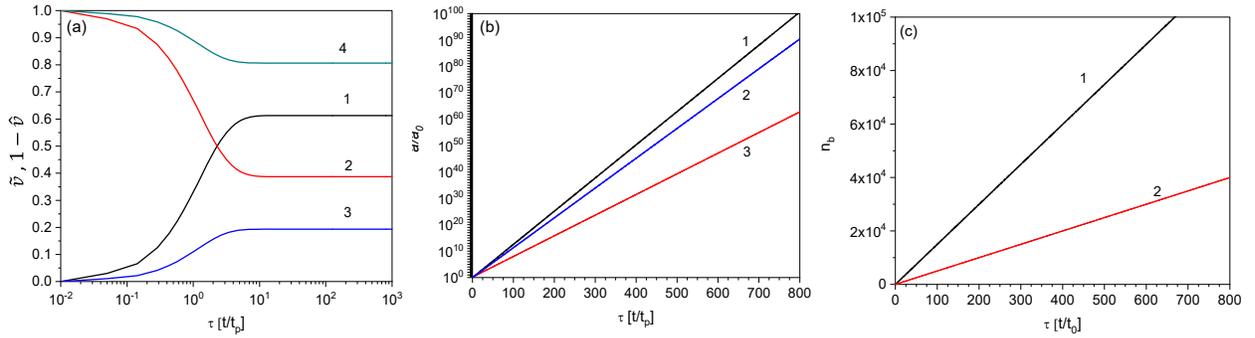

**Fig. 5.** Inflationary stage of expansion of the universe with discontinuities in density and finite energy density generated at the edges of the "bubble" of the physical vacuum (Section 4). The calculation parameters $\beta = 10^{-5}$, $\tilde{\rho}_V = 0.01$, $r_0 = 0.05$, $\xi_R = 0.04$, $\xi = 150$ and 50. (a) – dependences of the quantities $\tilde{v}$ (lines 1, 3) and $1 - \hat{v}$ (lines 2, 4) on time. Lines 1 and 2 correspond to $\xi = 150$, lines 3 and 4 correspond to $\xi = 50$. (b) – dependence of radius of the expanding universe on time. Line 1 corresponds to the case without generation of bubbles of real vacuum, lines 2 and 3 correspond to the case with generation of bubbles of real vacuum. Line 1 corresponds to $\xi = 150$, line 2 corresponds to $\xi = 50$. The vertical lines show the values at which the real vacuum displaces the false one and the radiation expansion of the universe begins. For $\tau \geq 5$ new discontinuities do not affect the value of $\hat{v}$ and correspondently on grows of $a$ (Fig.5,b). (c) – dependences of the number of raptures on time. Line 1 corresponds to $\xi = 150$ and line 2 corresponds to $\xi = 50$.

So far, we have considered the case of the formation of physical vacuum bubbles without taking into account the generation of radiation and massive particles at the boundaries of the metric discontinuities. Let us choose the following model of effective mass generation in a bubble. Let the mass density in the universe be proportional to the relative volume occupied by the real vacuum, $\tilde{\rho}_R = \tilde{v}\tilde{\rho}_{0,R}$, where $\tilde{\rho}_{0,R}$ is the density of matter in Planck units. Below (Section IV) it is shown that a physical vacuum bubble is a potential well for particles generated at the boundary between a false and physical vacuum – in other words, massive and massless particles do not spread throughout the entire volume of the universe, but are concentrated in "bubbles". Accordingly, equations (22) - (24) take the form:

$$\frac{1}{a}\frac{d^2a}{d\tau^2} = \frac{8\pi}{3}\tilde{\rho}_V(1-\tilde{v}) - \frac{8\pi}{3}\tilde{v}\tilde{\rho}_{0,R} = -\frac{8\pi}{3}\tilde{\rho}_V\big(1 - \tilde{v}(1+\xi_R)\big) \qquad (27)$$

$$\frac{dn}{d\tau} = \xi \cdot exp\left(-\frac{\beta}{\tilde{\rho}_V(1-\tilde{v}(1+\xi_R))}\right) \qquad (28)$$

$$\tilde{v} = \int_0^\tau \frac{a_0^3}{a(\tau')^3} \frac{dn}{d\tau'} \left(\frac{r_0}{a_0} + \frac{u(\tau-\tau')}{a_0}\right)^3 d\tau' \tag{29}$$

where, $u = dR_b/dt$ in (29) is the rate of expansion of the discontinuity boundary in units of the speed of light, independent of the rate of expansion of the universe. The question as to the reasons for the increase in the size of the "bubble" of the real vacuum, which is not associated with the general expansion of the universe, is considered in Section IV.

It is convenient to introduce the variable $\hat{v} = \tilde{v}(1 + \xi_R)$, transforming equations (27) - (29) to the form:

$$\frac{1}{a}\frac{d^2 a}{d\tau^2} = -\frac{8\pi}{3}\tilde{\rho}_V(1 - \hat{v}) \tag{30}$$

$$\frac{dn}{d\tau} = \xi \cdot exp\left(-\frac{\beta}{\tilde{\rho}_V(1-\hat{v})}\right) \tag{31}$$

$$\hat{v} = (1 + \xi_R)\xi \int_0^\tau \frac{a_0^3}{a(\tau')^3} \frac{dn}{d\tau'} \left(\frac{r_0}{a_0} + \frac{u(\tau-\tau')}{a_0}\right)^3 d\tau' \quad . \tag{32}$$

Equations (30) and (31) coincide with (22) and (23), respectively, and (32) at $u = 0$, up to the factor $(1 + \xi_R)$, coincides with (24).

After displacing the false vacuum by the physical $(1 - \tilde{v}) \ll 1$, the radiation expansion of the universe begins [1], which is described, in the adopted variables, by the equation:

$$\frac{1}{a}\frac{d^2 a}{d\tau^2} = -\frac{8\pi}{3}\xi_R\tilde{\rho}_V\frac{a_R^4}{a^4}, \tag{33}$$

where $a_R$ is the radius of the universe at which the false vacuum is almost completely replaced by the physical one.

Let us consider the case when at the initial moment of time $n = n_0$ discontinuities of the metric ("bubbles" of the physical vacuum) with radius $r_0 \ll a_0$ appear, after which their radii grow in accordance with equation (32). In this case, $\frac{dn}{d\tau'} = \delta(\tau')$, and equation (32) with $r_0 = 0$ takes the form:

$$\hat{v} = (1 + \xi_R)\xi n_0 \int_0^\tau \frac{a_0^3}{a(\tau')^3} \delta(\tau') \left(\frac{u(\tau-\tau')}{a_0}\right)^3 d\tau' = (1 + \xi_R)\xi n_0 \left(\frac{u\tau}{a_0}\right)^3 \tag{34}$$

Figure 6 shows the computed results for $\beta = 10^{-5}$, $\tilde{\rho}_V = 0.01$, $r_0 = 0.05$, $\xi_R = 0.04$, $u = 0.0005$, and $\xi = 150$ and 50. As seen from Fig. 6a, in a relatively short time on the order of $\tau \approx 3$, "bubbles" of physical vacuum are generated, which then expand until they displace the false one. Bubbles born after $\tau \approx 3$ practically do not contribute to the volume of physical vacuum.

Indeed, the main contribution to the displacement of the false vacuum by "bubbles" will be made by bubbles that have arisen in a time on the order of $\frac{1}{\gamma_V} = \left(\frac{8\pi}{3}\tilde{\rho}_V\right)^{-\frac{1}{2}} = 3.4$, the volume of which further increases as $\sim u^3 t^3$ until the relative volume of the physical vacuum displaces the false vacuum, i.e., $\hat{v} = 1$.

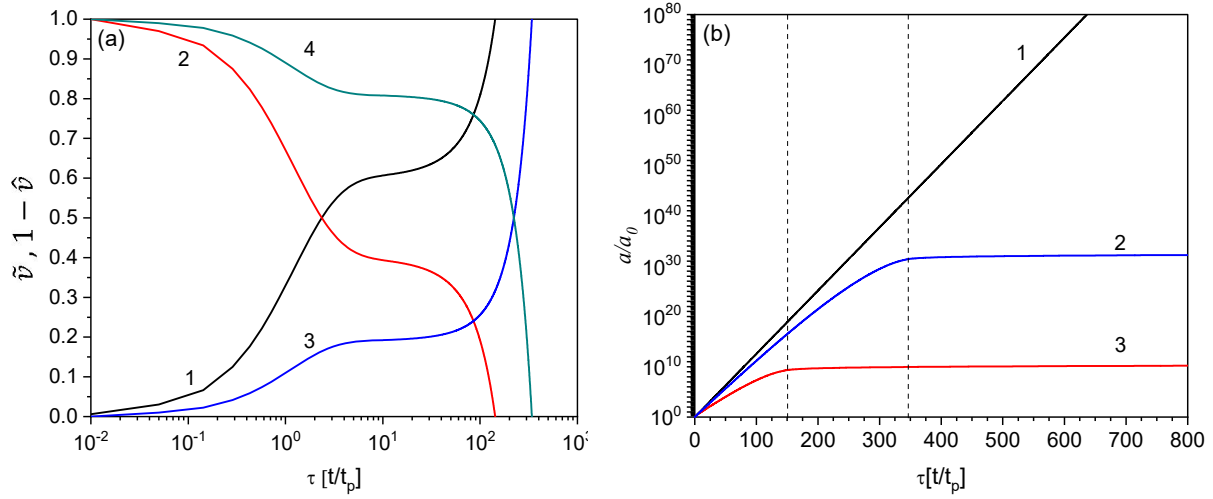

**Fig. 6.** Inflationary stage of expansion of the universe with discontinuities in density and finite energy density generated at the edges of the "bubble" of the physical vacuum (Section IV). The calculation parameters are $\beta = 10^{-5}$, $\tilde{\rho}_V = 0.01$, $r_0 = 0.05$, $\xi_R = 0.04$, $u = 0.0005$, and $\xi = 150$ and 50. (a) – dependences of the quantities $\tilde{v}$ (lines 1 and 3) and $1 - \hat{v}$ (lines 2 and 4) on time. Lines 1 and 2 correspond to $\xi = 150$, and lines 3 and 4 correspond to $\xi = 50$. (b) – dependence of radius of the expanding universe. Line 1 corresponds to the case without generation of bubbles of real vacuum, while lines 2 and 3 correspond to the case with generation of bubbles of real vacuum. Line 2 corresponds to $\xi = 150$, line 3 corresponds to $\xi = 50$. The vertical lines show the values at which the real vacuum displaces the false one and the radiation expansion of the universe begins. For $\tau \geq 5$ new discontinuities do not affect the value of $\hat{v}$ and correspondently on grows of $a$ (Fig.6,a).

Figure 7 shows the result of the calculation, with the same parameters as in Figure 6, now assuming $r_0 = 0$. The dotted lines in Fig. 7a shows a solution corresponding to (30).

In light of the above, the scenario of the initial stage of the expansion of the universe can be divided into three stages. At the first stage, over a fairly short period of time, a certain number of false vacuum breaks are generated in the universe and are evenly distributed. At the second stage, the resulting voids of real vacuum areas expand due to the generation of radiation energy at their boundaries, and this expansion rate exceeds the expansion rate of the universe because of the remaining false vacuum (see Section IV). At the third stage, the formed "bubbles" of real vacuum completely cover the entire volume of the universe, forming a foamy-like structure (radiation energy is concentrated in the bridges between the voids). At this stage, the generation of energy ends and the cooling of the universe and the process of nucleosynthesis begin, corresponding to the standard model of the Big Bang. As for dark energy and dark matter, in accordance with the described scenario, they should be concentrated in narrow, denser clusters, where most of the visible matter is concentrated, bordering on rarefied regions of the cosmic web.

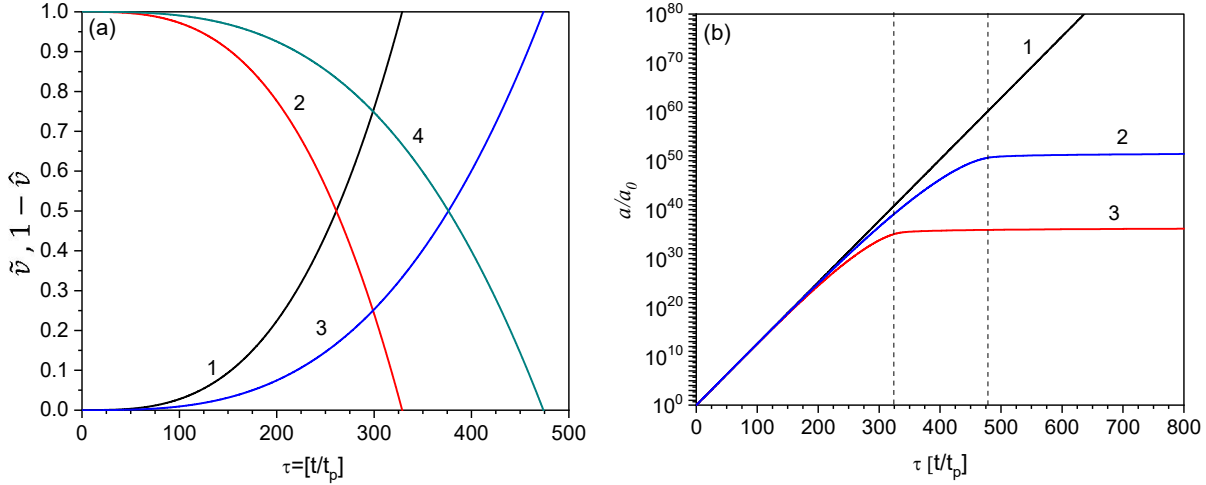

**Fig. 7.** Same as in Figure 6, but for $r_0 = 0$. Lines 1, 3 in Fig. 7,a practically coincide with the solution of equation (34). The vertical lines show the values at which the real vacuum displaces the false one and the radiation expansion of the universe begins.

## IV. The boundary of the real vacuum bubble

Consider the discontinuity of the space-time metric in the frame of reference associated with the center of a spherically symmetric discontinuity for a time much less than the characteristic time of the expansion of the universe $\tau_c \ll \frac{1}{\gamma_V} = \frac{1}{\sqrt{\frac{8\pi}{3}\tilde{\rho}_V}}$ schematically shown in Fig. 8.

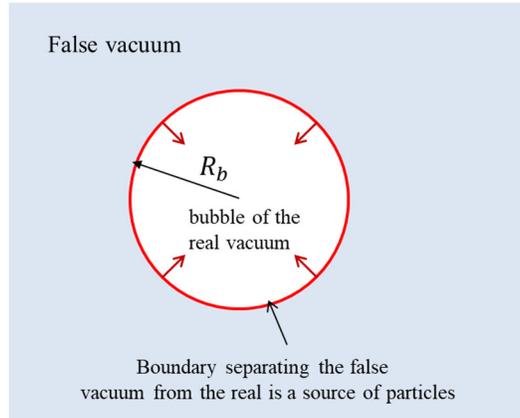

**Fig. 8.** A bubble of a physical vacuum in a false vacuum. The border separating the real vacuum from the false one is a source of energy (particles). The arrows show the direction of motion of the generated particles. They cannot leave the area of the bubble, due to the anti-gravitational forces acting on them from the side of the false vacuum.

In accordance with [49], the expression for the interval $ds$ in the centrally symmetric case in terms of the variables $t, r, \theta, \phi$ has the form:

$$ds^2 = e^\nu c^2 dt^2 - e^\lambda dr^2 - r^2(d\theta^2 + \sin^2\theta) \qquad (35)$$

In this case, Einstein's equations have the form [49]:

$$\frac{8\pi G}{c^4} T_0^0 - \frac{1}{c^2}\Lambda = -e^{-\lambda}\left(\frac{1}{r^2} - \frac{\lambda'}{r}\right) + \frac{1}{r^2} \tag{36}$$

$$\frac{8\pi G}{c^4} T_1^1 - \frac{1}{c^2}\Lambda = -e^{-\lambda}\left(\frac{\nu'}{r} + \frac{1}{r^2}\right) + \frac{1}{r^2} \tag{37}$$

$$\frac{8\pi G}{c^4} T_2^2 - \frac{1}{c^2}\Lambda = \frac{8\pi G}{c^4} T_3^3 - \frac{1}{c^2}\Lambda = -\frac{1}{2}e^{-\lambda}\left(\nu'' + \frac{\nu'^2}{2} + \frac{\nu'-\lambda'}{r} - \frac{\nu'\lambda'}{2}\right) + \frac{1}{2}e^{-\nu}\left(\ddot{\lambda} + \frac{\dot{\lambda}^2}{2} - \frac{\dot{\lambda}\dot{\nu}}{2}\right) \tag{38}$$

$$\frac{8\pi G}{c^4} T_0^1 = \frac{8\pi G}{c^4} T_1^0 = -e^{-\lambda}\frac{\dot{\lambda}}{r} \tag{39}$$

In (36)-(39) $T_k^i$ is the energy-momentum tensor. The dot above the letter denotes the derivative with respect to time, and the prime denotes the derivative with respect to the coordinate $r$.

In the area of the false vacuum ($r \geq R_b$), $R_b$ is the radius of the "bubble" of the physical vacuum, $T_k^i = 0$, $\Lambda \neq 0$, while in the area of the real vacuum ($r < R_b$) $T_k^i = 0$, $\Lambda = 0$. This implies that $\dot{\lambda} = 0$, and accordingly, that the second term on the right-hand side of (38) is equal to zero. Without going into the details of some straightforward calculations, we write out an expression for the $ds$ interval in the regions of false and real vacuum:

$$ds^2 = \begin{cases} r \geq R_b & \left(1 + \frac{\Lambda r^2}{3c^2}\left(1 - \frac{R_b^3}{r^3}\right)\right)c^2 dt^2 - r^2(\sin^2\theta\, d\phi^2 + d\theta^2) - \frac{1}{1+\frac{\Lambda r^2}{3c^2}\left(1-\frac{R_b^3}{r^3}\right)}dr^2 \\ r < R_b & c^2 dt^2 - r^2(\sin^2\theta\, d\phi^2 + d\theta^2) - dr^2 \end{cases} \tag{40}$$

The transition from a false vacuum to a real one requires special consideration that goes beyond the scope of the article, so we restrict ourselves to the assumption that this transition region exists and has a finite thickness $\delta r \ll r_0$.

Consider the forces acting on particles in the region of false and real vacua. Considering the gravitational field at the discontinuity to be independent of time, the force acting on a particle of mass $m$ in the centrally symmetric case, according to [49], is:

$$f = -\frac{mc^2}{\sqrt{1-\frac{v^2}{c^2}}}\frac{\partial}{\partial r}\left(\ln\sqrt{g_{00}}\right) \tag{41}$$

where $g_{00} = \left(1 + \frac{\Lambda r^2}{3c^2}\left(1 - \frac{R_b^3}{r^3}\right)\right)c^2$ (see (40) for $r \geq R_b$). From (41) we obtain the following expression for $f$:

$$f = -\frac{mc^2}{\sqrt{1-\frac{v^2}{c^2}}}\left(\frac{2}{3}\Lambda r\frac{\left(1+\frac{1R_b^3}{2r^3}\right)}{\left(1+\frac{\Lambda r^2}{3} - \frac{\Lambda R_b^3}{3r}\right)}\right) < 0. \tag{42}$$

The force at the border is directed towards the physical vacuum. This is to be expected since the cosmological constant $\Lambda$ acts as antigravity which "pushes" any particles out of the false vacuum.

From (42) it follows that the restoring force acting on the particle increases given a larger radius of the "bubble" of physical vacuum. At $r = R_b + 0$ (inside the false vacuum area), the force equals:

$$f = -\frac{mc^2}{\sqrt{1-\frac{v^2}{c^2}}} \Lambda R_b , \qquad (43)$$

and, for $\frac{\Lambda r^2}{3} \gg 1$,

$$f = -\frac{mc^2}{\sqrt{1-\frac{v^2}{c^2}}} \frac{2}{r} \qquad (44)$$

Let us qualitatively describe the process of matter generation at the boundary between physical and false vacuum using the example of the creation of a massive particle-antiparticle pair. Since the force acting on a particle with mass in the region of the false vacuum $r = R_b + 0$ is not equal to zero (formula (38)) and at the point $r = R_b - 0$ is equal to zero, then at the border of the false and real vacuum the tidal forces arise:

$$f_t = f_{+0} - f_{-0} \approx \frac{mc^2}{\sqrt{1-\frac{v^2}{c^2}}} \Lambda R_b . \qquad (45)$$

To create a particle-antiparticle pair, tidal forces must perform work equal to twice the rest mass of the particles at a length scale on the order of the De Broglie length of the particle (antiparticle).

This energy is taken from the negative energy of the false vacuum, defined by a non-zero $\Lambda$-term. Thus, the generation of a substance in a bubble of a physical vacuum is accompanied by the "melting" of a false vacuum, which determines the velocity $u$ in equation (29). The radius of the physical vacuum bubble grows with the expansion of the universe. Therefore, relative to the center of the bubble, the source of matter (energy) escapes with a velocity equal to the relativistic sum of the expansion velocity of the universe and the speed of "melting" of the false vacuum. Thus, at the border of each bubble of the physical vacuum, a thin layer of matter is formed, held by its own gravitational field and moving away from the center of the bubble.

**Conclusions**
1. A cavitation model of the inflation stage is proposed, in which it is assumed that at a certain value of negative pressure ($\Lambda$-term in Einstein's equations) metric discontinuities occur – regions of physical vacuum are formed where the $\Lambda$-term is equal to zero.
2. It is shown that the discontinuities of the metric during the transition from a false to a physical vacuum are an isotropizing factor, since they keep negative pressure in any part of the universe near the threshold of generation of discontinuities. Any local excess of negative pressure is quickly damped by an increase in the number of "bubbles" of physical vacuum.
3. It is shown that the isotropic and homogeneous distribution of the CMB and the visible distribution of matter is a natural consequence of metric discontinuities at the stage of inflationary expansion of the universe and does not depend on the rate of expansion of the universe.
4. It is shown that taking into account cavitation significantly slows down the inflation process.
5. The mechanism of generation of matter in the universe at the stage of inflationary expansion is indicated and it is shown that matter should be concentrated in the vicinity of boundaries of bubbles of the physical vacuum. A model of cavitation inflation is built taking into account energy generation. The transition from inflationary to radiation expansion is qualitatively analyzed.

6. The bubbles of physical vacuum can be the embryos of the large-scale cellular structure of the universe.
7. **Since the observed "voids" in the structure of the universe are the result of the evolution of the "bubbles" of the physical vacuum that arose at the initial stage of inflationary expansion, the observed size of the rarefied regions (voids) should decrease with increasing distance from an observer.**
8. **From the considered cavitation model it follows that dark energy and dark matter should be concentrated in the webs separating the "voids" from each other.**

**References**
[1] Ya.B. Zel'dovich, I.D. Novikow, "Relativistic Astrophysics: Volume 2: The structure and Evolution of the Universe" (University of Chicago Press, 1983)
[2] I.B. Arkhangelskaya, I.L. Rosenthal, A.D. Chernin, "Cosmology and physical vacuum", (Moscow: KomKniga, 2006) (in Russian)
[3] A.A Penzias, R.W. Wilson, A Measurement of Excess Antenna Temperature at 4080 Mc/s, Astrophysical Journal 142, 419 (1965)
[4] G. Gamov, Expanding Universe and the Origin of Elements, Phys. Rev. 70, 572 (1946)
[5] A.H. Guth, The Inflationary Universe: A possible Solution to the Horizon and Flatness Problems, Phys. Rev. D 23, 347 (1981)
[6] A. D. Linde, Nonsingular Regenerating Inflationary Universe, Print-82-0554 (CAMBRIDGE). Full text can be found at http://www.stanford.edu/~alinde/1982.pdf
[7] A. D. Linde, A New Inflationary Universe Scenario: A Possible Solution of the Horizon, Flatness, Homogeneity, Isotropy and Primordial Monopole Problems, Phys. Lett. B 108, 389 (1982)
[8] A. D. Linde, Inflation Can Break Symmetry in Susy, Phys. Lett. B 131, 330 (1983)
[9] A. D. Linde, Chaotic Inflation, Phys. Lett. B 129, 177 (1983)
[10] A. D. Linde, Eternally Existing Self-reproducing Chaotic Inflationary Universe, Phys. Lett. B 175, 395 (1986).
[11] A. Vilenkin, The Birth of Inflationary Universes, Phys. Rev. D 27, 2848 (1983)
[12] A. Vilenkin, Predictions from Quantum Cosmology, Phys. Rev. Lett. 74, 846 (1995)
[13] A. Vilenkin, Making Predictions in Eternally Inflating Universe, Phys. Rev. D 52, 3365 (1995)
[14] A. Vilenkin, "Many Worlds in One: The Search for Other Universes", (New York: Hill and Wang, 2007)
[15] W. Lerche, D. Lüst, A.N. Schellekens, Chiral Four-Dimensional Heterotic Strings from Self-Dual Lattices, Nucl. Phys. B 287, 477 (1987)
[16] M. Tegmark, A. Aguirre, M. Rees, F. Wilczek, Dimensionless Constants, Cosmology and other Dark Matters, Phys. Rev. D 73, 023505 (2006)
[17] B. Freivogel, Anthropic Explanation of the Dark Matter Abundance, J. Cosmology and Astroparticle Phys. 1003, 021 (2010)
[18] R. Bousso, L. Hall, Why Comparable? A Multiverse Explanation of the Dark Matter Baryon Coincidence, Phys. Rev. D 88, 063503 (2013)
[19] Y.B. Zel'dovich, A.A. Starobinsky, Quantum Creation of a Universe in a Nontrivial Topology, Sov. Astron. Lett. 10, 135 (1984)
[20] D.H. Coule, J. Martin, Quantum Cosmology and Open Universes, Phys. Rev. D 61, 063501 (2000)
[21] A.D. Sakharov, Cosmological Transitions With A Change In Metric Signature, Sov. Phys. JETP 60, 214 (1984) [Zh.Eksp.Teor.Fiz. 87, 375 (1984)]

[22] P. Byrne, "The Many Worlds of Hugh Everett III: Multiple Universes, Mutual Assured Destruction, and the Meltdown of a Nuclear Family" (NewYork: Oxford University Press, 2010)
[23] B. Greene, "The Hidden Reality: Parallel Universes and the Deep Laws of the Cosmos", First Vintage Books Edition (New York: Alfred A. Knopf, Inc. 2011)
[24] M. Tegmark, Parallel Universes, Scientific American. 288, 40 (2003)
[25] M. Kaku, "Parallel Worlds: A Journey Through Creation, Higher Dimensions, and the Future of the Cosmos" (New York: Random House, Inc. 2005).
[26] D. Deutsch, "The Ends of the Universe. The Fabric of Reality: The Science of Parallel Universe and Its Implications" (London: Penguin Press, 1997)
[27] R. Bousso, L. Susskind, L. Multiverse interpretation of quantum mechanics. Phys. Rev. D. 85 (4): 045007 (2012)
[28] Y. Nomura, Physical theories, eternal inflation, and the quantum universe, J. High Energy Phys. Article number: 63 (2011)
[29] P. Davies, "The Goldilocks Enigma: Why Is the Universe Just Right for Life?" (New York: A Mariner Book, Houghton Mifflin Co., 2008)
[30] P. Davies, A Brief History of the Multiverse. The New York Times (12 April 2003).
[31] P.J. Steinhardt, "Theories of Anything" WHAT SCIENTIFIC IDEA IS READY FOR RETIREMENT (9 March 2014). https://www.edge.org/responses/what-scientific-idea-is-ready-for-retirement
[32] A. Ijjas, A. Loeb, P. Steinhardt, Cosmic Inflation Theory Faces Challenges, Scientific American, 316 (2): 32 (February 2017)
[33] G.W. Gibbons, N. Turok, The Measure Problem in Cosmology. Phys. Rev. D. 77 (6): 063516 (2008)
[34] V. Mukhanov, Inflation without Selfreproduction, Fortschritte der Physik, 63 (1) 36 (2014)
[35] S. Perlmutter et al., Measurements of Ω and Λ from the First Seven supernovae at $z \geq 0.35$, Astrophysical J. 483, 565 (1997)
[36] S. Perlmutter et al., Measurements of Ω and Λ from 42 high redshift supernovae, Astrophys. J., 517, 565 (1999)
[37] A.G. Riess et al., Observational Evidence from Supernovae for an Accelerating Universe and a Cosmological Constant, The Astronomical J., 116, 1009 (1998)
[38] B.P. Schmidt, The High-Z Supernova Search: Measuring Cosmic Deceleration and Global Curvature of the Universe Using Type Ia Supernovae, The Astronomical Journal, 507, 46 (1998)
[39] M.N. Shneider, M. Pekker, "Liquid Dielectrics in an Inhomogeneous Pulsed Electric Field, Dynamics, Cavitation, and Related Phenomena (Second Edition)". (IOP Publishing, Temple Circus, Temple Way, Bristol, BS1 6HG, UK, 2019)
[40] M. Pekker, M.N. Shneider, Initial Stage Of Cavitation In Liquids and its Observation by Rayleigh Scattering, Fluid Dyn. Res. 49 035503 (2017)
[41] O M F R S L Rayleigh, On the Pressure Developed In A Liquid During the Collapse of a Spherical Cavity, Philos. Mag. 34, 94 (1917)
[42] Y.B. Zeldovich, Theory of Formation of a New Phase. Cavitation, Zh. Eksp. Teor. Fiz. 12, 525 (1942)
[43] J.C. Fisher, The Fracture of Liquids, J. Appl. Phys. 19, 1062 (1948)
[44] F. Caupin, E. Herbert, Cavitation in Water: a Review, C.R. Physique, 7, 1000 (2006)
[45] M.N. Shneider, M. Pekker, Cavitation in Dielectric Fluid in Inhomogeneous Pulsed Electric Field, J. Appl. Phys. 114, 214906 (2013)
[46] C.E. Campbell, R. Folk, E. Krotseheck, Critical Behavior of Liquid 4He at Negative Pressures, J. Low Temp. Phys., 105, 13 (1996)
[47] I.M. Lifshitz, Y. Kagan, Quantum Kinetics of Phase Transitions at Temperatures Close to Absolute, JETP, 35, 206 (1972)
[48] "CRC Handbook of Chemistry and Physics 2003–2004", D.R. Lide, (ed), 84th edn (Boca Raton, FL: CRC, 2004)

[49] L.D. Landau, E.M. Lifshitz, "The Classical Theory of Fields", 4th Edition, Course of Theoretical Physics, (Butterworth-Heinemann (Elsevier), 1975)